\documentclass[pra, showpacs, twocolumn, floatfix,superscriptaddress]{revtex4}
\usepackage{amsmath}
\usepackage{graphicx}
\usepackage{subfigure}
\usepackage{amsmath, amsfonts, amssymb, bm}

\allowdisplaybreaks[2]

\begin{document}
\title{Photon scattering from strongly driven atomic ensembles}
\author{Lu-ling \surname{Jin}}
\email{jinll@nwu.edu.cn}
\affiliation{Max-Planck Institut f\"ur Kernphysik, Saupfercheckweg 1,
D-69117 Heidelberg, Germany}
\affiliation{Department of Physics, Northwest University,  Xi'an 710069, Shaanxi, China}

\author{J\"{o}rg \surname{Evers}}
\email{joerg.evers@mpi-hd.mpg.de}
\affiliation{Max-Planck Institut f\"ur Kernphysik, Saupfercheckweg 1,
D-69117 Heidelberg, Germany}

\author{Mihai \surname{Macovei}}
\email{mihai.macovei@mpi-hd.mpg.de}
\affiliation{Max-Planck Institut f\"ur Kernphysik, Saupfercheckweg 1,
D-69117 Heidelberg, Germany}

\date{\today}
\begin{abstract}
The second order correlation function for light emitted from a strongly and near-resonantly driven dilute cloud of atoms is discussed. Because of the strong driving, the fluorescence spectrum separates into distinct peaks, for which the spectral properties can be defined individually. It is shown that the second-order correlations for various combinations of photons from different spectral lines exhibit bunching together with super- or sub-Poissonian photon statistics, tunable by the choice of the detector positions. Additionally, a Cauchy-Schwarz inequality is violated for photons emitted from particular spectral bands. The emitted light intensity is proportional to the square of the number of particles, and thus can potentially be intense. Three different averaging procedures to model ensemble disorder are compared.
\end{abstract}
\pacs{42.25.Hz, 42.50.Hz, 42.50.Ar, 42.50.Dv}
\maketitle

\section{Introduction}
The characterization of light is an ubiquitous problem, and a convenient formalism for this is the use of correlation functions. However it is well known that it is not possible to sufficiently distinguish the nature of a given light source from only the first-order correlation function \cite{RJG}. In particular, the quantum properties of light cannot be extracted from the first order correlation function. This motivated the study of second-order correlations, initiated by the intensity-correlation experiments conducted by Hanbury-Brown and Twiss \cite{HBT1}. Subsequently, second-order correlation measurements have found applications in many fields of modern physics \cite{mp} such as astronomy \cite{HBT2}, optics \cite{ar}, high-energy physics \cite{hep}, condensed matter physics \cite{cmp,dif} and atomic physics \cite{ap}.

Next to the characterization of a physical system via higher-order correlations, it has also been suggested to use strongly correlated particles or even entangled particles as input to a system for various applications, with the most obvious examples of quantum computation, quantum communication and quantum information processing. This prompts the question for efficient sources of correlated photons. A standard method for generating entangled photons is the parametric down-conversion processes \cite{kl}. Alternatively, entangled photons can be generated in four-wave mixing \cite{fwm1,fwm2} or electromagnetically induced transparency \cite{eit} processes. Further, an experiment on photon anti-bunching in phase-matched multi-atom resonance fluorescence was reported in \cite{ab_m}. Non-classical photon pairs for scalable quantum communication with atomic ensembles and ultraviolet entangled photons in a semiconductor were generated in \cite{kuz} and \cite{semi}, respectively. An atomic memory for correlated photon states was demonstrated in \cite{lk}, and a heralded entanglement source was realized  in~\cite{her}. However, despite the large variety of sources, in many cases applications are restricted by a limited production rate for correlated photons, and by the lack of sources with tunable correlations.

\begin{figure}[b]
\includegraphics[width=7.7cm]{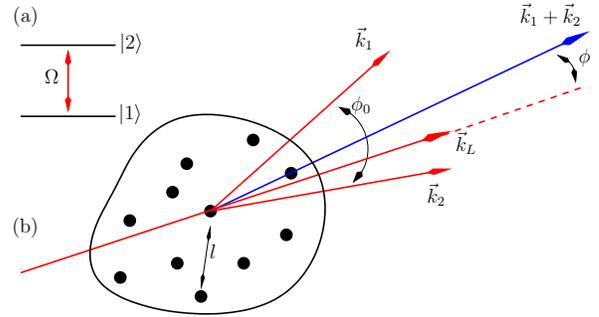}
\caption{\label{fig1}(Color online) Schematic setup of a dilute atomic ensemble pumped with
a coherent field with wave-vector $\vec k_{L}$. (a) shows the energy levels
of each of the ensemble particles, and the interaction with the strong coherent light
with coupling strength $\Omega$. (b) depicts the ensemble with typical inter-particle
distance length scale denoted by $l$ with $l \gg 2\pi/k_{L}$. We consider the case
of photon pair emission in forward direction and denote the angle between the two emitted
photons with wave-vectors $\vec k_{1}$ and $\vec k_{2}$ as $\phi_{0}$, and
 $\vec k_{1} + \vec k_{2} \approx 2\vec k_{L}$. The direction
of the emission cone defined by $\vec k_{1}$ and $\vec k_{2}$ is characterized by
the angle $\phi$ between $\vec k_{1} + \vec k_{2}$ and $\vec k_{L}$.
}
\end{figure}

Interestingly, it has been shown that disordered media can sometimes exhibit surprising correlation or coherence effects. For example, speckle patterns in the quantum correlations within light scattered by a disordered medium were observed in~\cite{SP}. A cooperative radiation force in the presence of disorder was observed as well~\cite{ks}. An experimental investigation of the origin of  disorder in parametrically excited waves on a fluid surface (Faraday waves) was performed in~\cite{fr}. Instabilities of waves in nonlinear disordered media and subwavelength spatial correlations in near-field speckle patterns were discussed in~\cite{scr,carm}. Correlation and coherence can also be observed in the context of localization~\cite{ML}. Furthermore, coherent  backscattering was investigated in~\cite{bks}, and random lasing in~\cite{rl,rl1,rl2}.

In this paper, we study the generation of correlated light from a disordered many-particle system, and in particular focus on the case of a strong near-resonant driving of the ensemble. Due to the strong driving, the fluorescence spectrum dissolves into distinct lines, which can be characterized separately. We then investigate second-order correlation functions of light scattered into the various spectral lines. The disordered ensemble is modelled by calculating the emitted light for a fixed orientation between the atoms, followed by a suitable configuration averaging. For this, we compare three different averaging procedures. We find that the second-order correlation functions typically exhibit strong correlations if detected close to the forward scattering direction, whereas the corresponding first-order correlation functions are isotropic. In particular, bunching together with super- or sub-Poissonian photon statistics can be achieved for different combinations of spectral lines, and in some cases the statistics is tunable by the choice of the detector positions. We additionally demonstrate that light emitted from certain spectral bands violates a Cauchy-Schwarz inequality. Since the unnormalized second-order correlation functions of the emitted light scales with the number of particle in the ensemble squared, intense beams of correlated photons can be generated.

\section{Theoretical considerations}
\subsection{The model}
We proceed by investigating an atomic sample of
arbitrary shape and of characteristic size $d$, consisting of distinguishable non-overlapping
two-level particles. We assume conditions such that multiple scattering effects can be neglected, i.e., we consider the single scattering regime in which the photon mean free path length is of the order of the sample size or larger. Related, we also restrict the typical inter-particle separation $l$ to satisfy
\begin{align}
\label{conditions}
\lambda_{L} \ll l\ll d\,,
\end{align}
with $d/c < \tau_{s}$, where $\tau_{s}$
is the spontaneous decay time. All particles have identical atomic transition
frequencies $\omega_{0}$, and are localized at random positions
$\vec{r}_j$ with  $j\in\{1,2, \cdots, N\}$. We define the interparticle separation vectors
as $\vec{r}_{ij} = \vec{r}_i - \vec{r}_j$. The external laser field has frequency
$\omega_{L}= ck_{L}=2\pi c/\lambda_{L}$, wave vector $\vec k_{L}$ and
wavelength $\lambda_{L}$ (see Fig.~\ref{fig1}).

Under the action of the laser field, the system is best described in a suitable dressed state picture. In electric dipole and rotating wave approximations, the system Hamiltonian can be written as $H = H_{0} + H_{I}$, where \cite{melk}
\begin{subequations}
\begin{align}
H_{0} &= \sum_{k}\hbar(\omega_{k}-\omega_{L})a^{\dagger}_{k}a_{k} +
\sum^{N}_{j=1}\hbar \tilde \Omega_{j}R_{zj}\,,\\
H_{I} &= i\sum_{k}\sum^{N}_{j=1}(\vec g_{k}\cdot \vec d_{j})
\bigl \{a^{\dagger}_{k}S^{-}_{j}e^{-i(\vec k-\vec k_{L})\cdot \vec r_{j}}
-{\rm H.c.}\bigr\}
\label{HI}\,,\\
S^{-}_{j} &= \frac{R_{zj}}{2}\sin{2\theta_{j}} -R^{(j)}_{21}\sin^{2}{\theta_{j}}
+ R^{(j)}_{12}\cos^{2}{\theta_{j}}\label{ops}\,.
\end{align}
\end{subequations}
Here, $H_{0}$ represents the Hamiltonian of the
free electromagnetic field (EMF) and free dressed atomic subsystems, respectively,
while $H_{I}$ accounts for the interaction of the laser-dressed atoms with the EMF.
$a_{k}$ and $a^{\dagger}_{k}$ are the field annihilation and creation operators
obeying the standard commutation relations for bosons. The atomic operators
$R^{(j)}_{\alpha \beta}=|\tilde\alpha\rangle_{j} {}_{j}\langle \tilde \beta|$ describe
the transitions between the dressed states $|\tilde \beta \rangle_{j}$ and
$|\tilde \alpha \rangle_{j}$ in atom $j$ for $\alpha \not=\beta$ and dressed-state
populations for $\alpha=\beta$, and satisfy the commutation relations of the su(2) algebra.
The dressed states $|\tilde \alpha \rangle_{j}$ entering the operators $R^{(j)}_{\alpha \beta}$
can be represented through the bare states $|\alpha \rangle_{j}$ via the transformations
\begin{subequations}
\begin{align}
|1\rangle_{j} &= \sin{\theta}|\tilde 2\rangle_{j}+\cos{\theta}|\tilde 1\rangle_{j}\,,\\
|2\rangle_{j} &= \cos{\theta}|\tilde 2\rangle_{j}-\sin{\theta}
|\tilde 1\rangle_{j}\,.
\end{align}
\end{subequations}
We further defined
\begin{align}
R_{zj}=|\tilde 2\rangle_j {}_j\langle \tilde 2| - |\tilde 1\rangle_j {}_j\langle \tilde 1|\,,
\end{align}
which is the difference of the
upper and lower dressed state population. Further,
\begin{align}
\tilde \Omega=\tilde \Omega_{j}= \sqrt{\Omega^{2}+(\Delta/2)^{2}}
\end{align}
is the generalized Rabi frequency, with
$2\Omega=(\vec d\cdot \vec E_{L})/\hbar$. Here, $\vec E_{L}$ is the electric laser
field strength, and $\vec d \equiv \vec d_{j}$ is the transition dipole matrix element.
The detuning $\Delta = \omega_{0} - \omega_{L}$ is characterized by
$\cot{2\theta}=\Delta/(2\Omega)$.

\subsection{Spectral decomposition}
In the following we make use of the fact that the resonance fluorescence spectrum of the light scattered by an atomic system pumped by a strong near-resonant driving field splits up into distinct lines, which are known as the Mollow spectrum~\cite{Mollow}. The condition for well-resolved spectral lines is $\tilde \Omega \gg \gamma$, with $\gamma = 1/\tau_s$ being the single-atom spontaneous decay rate.
In this limit of strong driving, it is reasonable to define optical properties
for each of the spectral lines separately. Thus, it follows from the
interaction Hamiltonian Eq.~(\ref{HI}) that the different operators
in Eq.~(\ref{ops}) can be considered as sources of individual spectral lines.
In particular, in the following we will denote light originating from
$R_{zj}\sin(2\theta)/2$ as the central spectral component indicated by $C$,
and $R^{(j)}_{21}\cos^{2}{\theta}$ and $R^{(j)}_{12}\sin^{2}{\theta}$
as the right (R) and left (L) spectral sideband components. These sidebands
are emitted at  frequencies $\omega_{+} \equiv \omega_{L} + 2\tilde \Omega$
and $\omega_{-} \equiv \omega_{L}-2\tilde \Omega$, respectively
\cite{kilin,Tann}.
In what follows, we shall use this decomposition to
investigate the properties of the scattered light.

\subsection{First order correlations}
In order to calculate the light scattered by the cloud of two-level scatterers,
we assume detection in the far-zone limit, that is, the linear dimension of the atomic
system $d$ is much smaller than the distances between the cloud's center-of-mass
and the detector at $\vec R$.
The intensity of the scattered light can be calculated from the first order photon correlation function as
\begin{align}
\label{intensities}
I_{m}(\vec R) &= \langle a^{\dagger}_{m}(\vec{R}) a_{m}(\vec{R}) \rangle \,,
\end{align}
where $a_m$ is a photon operator for the $m$th spectral band with $m \in \{C,~R,~L\}$.
If was shown in~\cite{melk} that these intensities of the different
spectral lines can be evaluated as
\begin{subequations}
\begin{align}
I_{C}(\vec R)& = \frac{1}{4} \sum^{N}_{j,i=1}\Psi_{R}(\vec r_{ji},\omega_{L})
\: \langle R_{zj}R_{zi}\rangle\:  \sin^{2}(2\theta)\,,\\
I_{L}(\vec R)& = \sum^{N}_{j,i=1}\Psi_{R}(\vec r_{ji},\omega_{-})
\: \langle R^{(j)}_{12}R^{(i)}_{21}\rangle \: \sin^{4}{\theta}\,,\\
I_{R}(\vec R)& = \sum^{N}_{j,i=1}\Psi_{R}(\vec r_{ji}, \omega_{+})
\: \langle R^{(j)}_{21}R^{(i)}_{12}\rangle\:  \cos^{4}{\theta}\,.
\end{align}
\end{subequations}
Here, $\Psi_{R}(\vec r_{ji},\omega)=\Psi_{R}(\omega)
\exp{[i(\vec k-\vec k_{L})\vec r_{ji}]}$ for a scattered photon of
wave-vector $\vec k$. $\Psi_{R}(\omega)$ depends on the atom-environment coupling and in general
is a function of frequency with $R=|\vec R| \gg k^{-1}_{L}$.
As we focus on large inter-particle separations [see Eq.~(\ref{conditions})], there is only a negligible direct coupling between the atoms, and thus the collective correlators $\langle R^{(i)}_{12}R^{(j)}_{21}\rangle$ describing the vacuum-mediated interactions among the emitters $i$ and $j$ decouple for $i\not=j$, that is,
\begin{align}
\langle R^{(i)}_{12}R^{(j)}_{21}\rangle \approx \langle R^{(i)}_{12}\rangle
\langle R^{(j)}_{21}\rangle \qquad (i\neq j)\,.
\end{align}
In the strong driving field case,  the atomic variable
$\langle R^{(j)}_{12}\rangle$ scales as $\gamma/\tilde \Omega$ and therefore can
be neglected in a secular approximation. As a
consequence, the intensities of the spectral side-bands are proportional to the number
of scatters in the ensemble, i.e., $\{I_{L},I_{R}\} \propto N$. Near the resonance ($\theta=\pi/4$), the intensity of the central spectral line is also proportional
to $N$. In what follows, we will restrict the analysis to the resonant driving field case in which the
intensities of all spectral lines are proportional to the number of scatters $N$. In
this case, the intensities of spontaneously scattered photons are distributed uniformly
over the whole $4\pi$ solid angle.
Note that this is in contrast to the weak pumping case $\Omega < \gamma$, in which
the first-order correlation function  does show directional behavior as
\begin{align}
I(\vec R)&=\Psi_{R}\: \left \{N \left (\frac 12+\langle S_{z}\rangle \right ) \right .\nonumber\\
&\left. + N(N-1)\:|\langle S^{+}\rangle|^{2} \:\cos[(\vec k-\vec k_{L})\vec r_{ji}] \right\}\,, \label{wf}
\end{align}
where
\begin{subequations}
\begin{align}
\langle S_{z}\rangle & =-\frac{\gamma^{2}+\Delta^{2}}{2\bigl(\gamma^{2}+\Delta^{2}+\Omega^{2}\bigr)}\,,\\
\langle S^{+}\rangle&=\frac{i\Omega(\gamma^{2}+\Delta^{2})}{(\gamma-i\Delta)\bigl(\gamma^{2}+\Delta^{2}+\Omega^{2}\bigr)}\,.
\end{align}
\end{subequations}

\subsection{Second order correlations}
Our main observable in the following will be the the normalized second-order correlation functions $g^{(2)}(\tau)$, which is measured by two detectors positioned at $\vec{R}_1$ and $\vec{R}_2$. It  is defined as
\begin{eqnarray}
g^{(2)}_{mn}(\tau, \vec R_{1}, \vec R_{2})=
\frac{G^{(2)}_{mn}(\tau,\vec R_{1},\vec R_{2})}
{I_{m}(\vec R_{1})I_{n}(\vec R_{2})}, \label{crf}
\end{eqnarray}
i.e., as the unnormalized second order correlation function
\begin{align}
G^{(2)}_{mn}(\tau,\vec R_{1},\vec R_{2})=&\langle a^{\dagger}_{m}(\vec R_{1})
a^{\dagger}_{n}(\tau,\vec R_{2})\nonumber\\
&\times a_{n}(\tau,\vec R_{2})a_{m}(\vec R_{1})\rangle\,,
\end{align}
normalized to the intensities $I_{m}(\vec R_{1})$ and $I_{n}(\vec R_{2})$
defined in Eq.~(\ref{intensities}).
The quantity $g^{(2)}_{mn}$ for $\{m,n\} \in \{C,~R,~L\}$ can be interpreted
as a measure for the probability of detecting one photon emitted in mode
$m$ and another photon emitted in mode $n$ with time-delay $\tau$.
Particularly, $g^{(2)}(\tau=0)$ describes the photon statistics (e.g., sub/super-Poissonian), whereas $g^{(2)}(\tau \neq 0)$ indicates photon bunching or antibunching.

To calculate the correlation function, we assume laser driving on
resonance ($\theta = \pi/4$), and a large atomic ensembles ($N \gg 1$) such that the secular approximation is valid. We also assume that all possible pairs of atoms contribute equally to the second-order correlation functions. This assumption is
valid as long as the angle between the wave vectors of the incident laser and the scattered photons is small, that is $\{\phi,\phi_{0}\}$ should be of order of few
degrees  (see Fig.~\ref{fig1}). Finally, we for the moment consider a single interparticle distance vector $\vec{r}_{ji}$ for all pairs only, but this restriction will be relaxed later on. Based on these assumptions,  we found that the correlation and cross-correlation functions of photons scattered into the different spectral bands can be represented as
\begin{subequations}
\label{cors}
\begin{align}
g^{(2)}_{CC}(\tau,\vec R_{1},\vec R_{2}) &= 1 + 2\cos(\delta_{1})\cos(\delta_{2})e^{-2\gamma\tau}\,,\nonumber \\
&= 1 + \left [ \cos(\delta_{+})+ \cos(\delta_{-}) \right ] \,e^{-2\gamma\tau}\,,\\
g^{(2)}_{LL}(\tau,\vec R_{1},\vec R_{2})&=g^{(2)}_{RR}(\tau,\vec R_{1},\vec R_{2})
\nonumber \\
&=1 + \cos(\delta_{-})e^{-3\gamma\tau}\,,\\
g^{(2)}_{LR}(\tau,\vec R_{1},\vec R_{2})&= g^{(2)}_{RL}(\tau,\vec R_{1},\vec R_{2})
\nonumber\\
&=1 + \cos(\delta_{+})e^{-3\gamma\tau}\,,\\
g^{(2)}_{CX}(\tau,\vec R_{1},\vec R_{2}) &= g^{(2)}_{XC}(\tau,\vec R_{1},\vec R_{2})
\nonumber\\
&=1\, \qquad {\rm for}~ X \in \{L,R\}.
\end{align}
\end{subequations}
Here,
\begin{subequations}
\begin{align}
\label{delta}
\delta_{s}&=(\vec k_{s}-\vec k_{L})\vec r_{ji}\,,\\
\delta_{+}&=\delta_{1}+\delta_{2}\,,\\
\delta_{-}&=\delta_{1}-\delta_{2}\,,
\end{align}
\end{subequations}
 with $\vec k_{s}$ being
the wave-vector of the photon $s$ scattered in direction
$\vec R_{s}$ ($s\in\{1,2\})$.
Note that these results differ from the corresponding expressions for a single pair of atoms or a regular structure of atoms~\cite{melk}. The reason is that for a single pair, one chooses $N=2$, whereas in deriving Eqs.~(\ref{cors}) the assumption $N\gg1$ was used such that terms proportional to $1/N$ can be neglected.

\subsection{Ensemble averaging}
To estimate the signal obtained from a cloud of randomly distributed particles, the correlation functions $g_{XY}^{(2)}(\tau, \vec{R}_1, \vec{R}_2)$ have to be averaged over the different interatomic distance vectors $\vec{r}_{ij}$ in the cloud. Since in general it can be expected that the averaging procedure affects the result,  we compare three averaging procedures.
\subsubsection{\label{avg1}Averaging over a spherical shell}
The first averaging procedure we consider has been suggested in previous works on scattering from dilute gases, see, e.g.,~\cite{shat}. It consists of (i) an isotropic average over the relative orientation ${\bf n}$ of the atoms over the unit sphere, followed by (ii) an average of the inter-atomic distance $r_{ji}=|\vec r_{ji}|$ over an interval of  order of the laser wave-length, around their typical distance $l$ :
\begin{align}
\label{eqavg1}
\langle \cdots \rangle_{\rm conf}=
\frac{k_{L}}{(4\pi)^2}\:\int^{l+2\pi/k_{L}}_{{l-2\pi/k_{L}}}dr_{ji}\int d\Omega_{\bf n} \cdots\,.
\end{align}
It should be noted that the restriction to the radial averaging over a range of order of the wave length is not a priori justified; it is however motivated by the fact that the distance distribution of particles in a gas is peaked around a mean value, even though the distribution generally is broad. In the numerical examples below, we use $l=20\lambda_L$.

\subsubsection{\label{avg2}Averaging over a spherical volume}
This method is a modification of the first method in which we change the averaging over the radial coordinate to range from $0$ to twice the volume radius $R$:
\begin{align}
\label{eqavg2}
\langle \cdots \rangle_{\rm conf}=
\frac{1}{8\pi R}\:\int^{2R}_{{0}}dr_{ji}\int d\Omega_{\bf n} \cdots\,.
\end{align}
This averaging procedure marks the opposite extreme of the averaging in Sec.~\ref{avg1} in that a constant distance distribution over all possible distances is assumed.  In the numerical examples below, we use $R=100\lambda_L$.

\subsubsection{\label{avg3}Numerical sampling}
The first two averaging methods made assumptions on the distribution of distances between the atoms, which may not be fulfilled in a gas of atoms. To investigate the effect of this further, in a third averaging procedure, we randomly place a number of atoms $N$ in a cubic volume of side length $2R$ and then calculate the quantities in Eqs.~(\ref{cors}) for this sample of atoms. In the numerical examples below, we use $R=100\lambda_L$ and $N=300$.

\section{Results}

As a first result, we note from Eqs.~(\ref{cors}) that the second order correlation function can exceed unity in several cases together with bunching, which means that there is an enhanced probability to generate photons in pairs. For example, the second-order correlation function
for the central-band photons $g^{(2)}_{CC}$ has maxima when
$\vec k_{1}+\vec k_{2}=2\vec k_{L}$
[see Eq.~\ref{cors}(a)], i.e., in forward direction.
Similar results are observed for the sideband spectral lines. For example, pairs of photons can be generated in which one photon is emitted in the left spectral band, whereas the other is emitted from the right spectral sideband, see Eq.~\ref{cors}(c).
In contrast, photon pairs originating from the same sideband (left or right) are most likely for $\vec k_{1}=\vec k_{2}$ independent of the direction of the wave vector and thus do not show directional behavior, see Eq.~\ref{cors}(b).
Finally, there is no enhanced probability for detecting pairs of photons in which one
photon is generated in the central band and one in either of the sidebands, and the corresponding correlation function does not exhibit any directionality, see Eq.~\ref{cors}(d).

\begin{figure}[t]
\centering
\includegraphics[width=0.95\columnwidth]{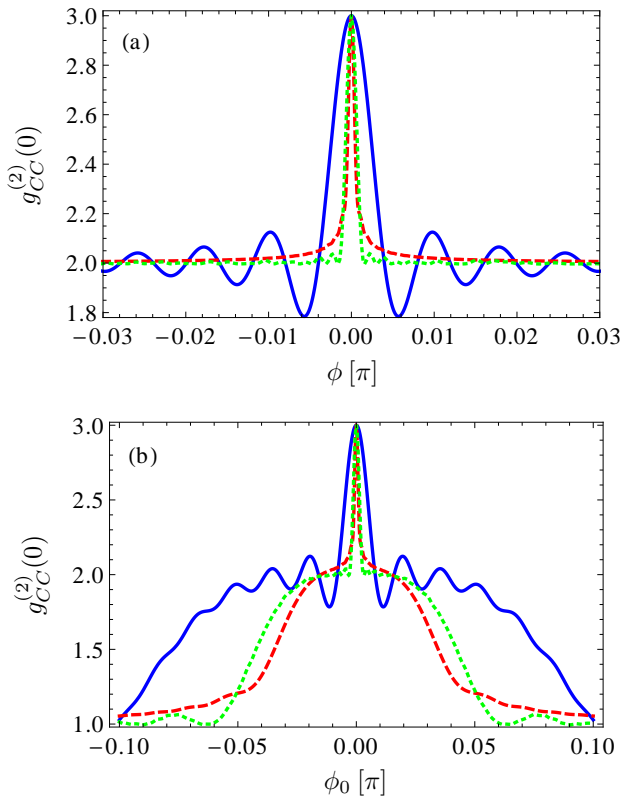}
\caption{\label{fig2}(Color online) Normalized and configuration averaged second-order
correlation function $g_{CC}^{(2)}(0)$ between two photons
emitted from the central spectral band. 
In (a), the correlation function is shown  for pairs of  photons emitted in the same direction ($\phi_{0}=0$, see Fig.~\ref{fig1}), and plotted as a function of the emission direction  $\phi$. 
In (b), the correlation function is plotted as a function of the opening angle $\phi_{0}$ between the two photons, which corresponds to the case of two distinct detectors. The two photons are measured at positions symmetric with respect to the incident laser field direction, i.e., $\phi=0$.
The blue solid line shows averaging over a spherical shell, the dashed red line averaging over a spherical volume, and the green dotted line the numerical sampling.
}
\end{figure}

\begin{figure}[t]
\centering
\includegraphics[width=0.95\columnwidth]{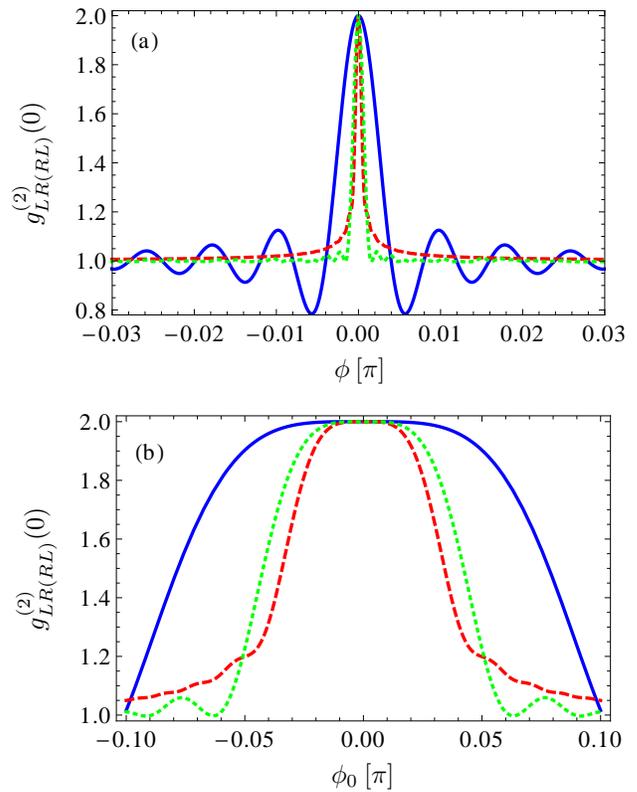}
\caption{\label{fig3}(Color online) Same as Fig.~\ref{fig2}, except that the normalized and configuration averaged second-order correlation functions $g_{LR(RL)}^{(2)}(0)$ between one photon emitted in the left and one photon emitted in the right sideband is shown.}
\end{figure}

Regarding the intensity, the unnormalized second-order correlation
functions are proportional to the number of particles in the ensemble squared, i.e.
$G^{(2)}(0) \propto N^{2}$.  

The obtained results for the normalized second-order correlation function $g_{CC}^{(2)}(0)$ for photons emitted in the central band after the configuration average are shown in Fig.~\ref{fig2}. It can be seen that even after the averaging, the correlation function exhibits a sharp peak around direction of the the incident laser wave-vector $\vec k_{L}$, indicating super-bunching. Fig.~\ref{fig2}(a) shows the case with a single two-photon detector ($\phi_{0}=0$) for different emission directions $\phi$. Figure~\ref{fig2}(b) depicts the same correlation function but with two single-photon detectors placed symmetrically with respect to the laser wave-vector $\vec k_{L}$ direction given by  $\phi=0$. 
Note that the correlation functions for two photons emitted both from the left or both from the right spectral sideband do not show a directionality in space.

Fig.~\ref{fig3} shows the corresponding results for the two photon cross correlation with one photon emitted in the left, and one in the right spectral sideband. Again, we find maxima at $\phi_{0}=0$ for two individual detectors placed symmetrically around the incident laser direction. But in contrast to the central band correlation function, this maximum is not peaked, but rather broad. If a two-photon detector is used, a narrow maximum is observed in the forward direction, see Fig.~\ref{fig2}(a). Interestingly, in this case, depending on the precise positioning around the forward direction and on the averaging procedure, also sub-Poissonian photon-statistics can occur.

\begin{figure}[t]
\centering
\includegraphics[width=0.95\columnwidth]{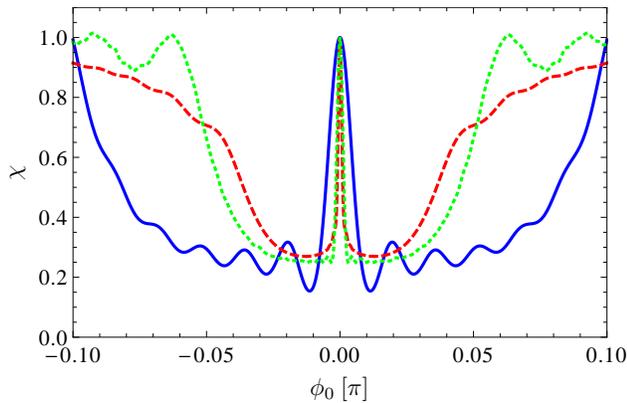}
\caption{\label{fig4}(Color online) The Cauchy-Schwarz inequality $\chi$ characterizing the
cross-correlations of photon pairs with one photon emitted from each sideband. The photons are
detected symmetrically around the incident laser field direction ($\phi=0$), and the result
is shown as a function of the angle between the wave vectors of the two emitted photons $\phi_{0}$. The Cauchy-Schwarz inequality is violated for $\chi<1$. The curves are as in Fig.~\ref{fig2}.}
\end{figure}

We further define the Cauchy-Schwarz parameter
\begin{align}
\chi=\frac{g^{(2)}_{LL}(0)\: g^{(2)}_{RR}(0)}{\left [g^{(2)}_{LR}(0)\right]^{2}}
=\frac{g^{(2)}_{LL}(0)\: g^{(2)}_{RR}(0)}{\left [g^{(2)}_{RL}(0)\right]^{2}}\,,
\end{align}
which relates the
correlation between photons emitted into individual modes to the cross-correlation
between photons emitted into two different modes \cite{CSI}. If $\chi <1$ the
Cauchy-Schwarz inequality is violated. Figure \ref{fig4} shows violation of the
Cauchy-Schwarz inequality for photons scattered into the  side-bands.

\section{Discussion and Summary}
An interpretation of the results obtained in the previous section can be found in the semiclassical dressed-state
picture. Suppose one laser photon is absorbed and the atom is in the dressed-state
$|\tilde 2\rangle$. Then one option will be a spontaneously emitted photon on the
$|\tilde 2\rangle \to |\tilde 2\rangle$ transition followed by a further absorption
of a laser photon during time-interval $\tilde \Omega^{-1}$ and subsequently decay
on the same dressed-state transition. Similar effect occurs on the dressed-state
transition $|\tilde 1\rangle \leftrightarrow |\tilde 1\rangle$ when initially the
emitter is in the $|\tilde 1\rangle$ dressed-state, and both processes contribute to
the central-band scattering where $\vec k_{1}+\vec k_{2} \approx 2\vec k_{L}$. Another
option will be a spontaneously emitted photon on the $|\tilde 2\rangle \to |\tilde 1\rangle$
dressed-state transition followed by an absorption of a laser photon on the
$|\tilde 1\rangle \to |\tilde 1\rangle$ transition and further spontaneously decay on
the $|\tilde 1\rangle \to |\tilde 2\rangle$ dressed-state transition. Together with the
reverse process where a spontaneous emission occurs on the $|\tilde 1\rangle \to |\tilde 2\rangle$
transition (if initially the particle is in the $|\tilde 1\rangle$ dressed-state) followed by
a laser absorption on the $|\tilde 2\rangle \to |\tilde 2\rangle$ transition and a subsequently
decay on the $|\tilde 2\rangle \to |\tilde 1\rangle$ dressed-state transition, these effects
describe photon correlations among the right and left side-bands, or vice versa and, again,
$\vec k_{1}+\vec k_{2} \approx 2\vec k_{L}$, and therefore, we have a directional-dependent photon
distribution. In contrast, all other scattering processes, i.e., two-photon spontaneously
emission on the same side-band or cross-correlations involving a side-band photon and a central
one do not fulfill the relation $\vec k_{1}+\vec k_{2} \approx 2\vec k_{L}$, that is
$\vec k_{1}+\vec k_{2} \not= 2\vec k_{L}$, and thus, in these processes the photon distribution
will be uniformly in space after a configuration averaging.

Comparing the results for the three averaging procedures in Figs.~\ref{fig2}-\ref{fig4}, it can be seen that the averaging strategy can substantially affect the result quantitatively, even though the qualitative features of the obtained results remain similar. In particular the averaging over a spherical volume and the numerical sampling in a cubic volume of comparable size agree reasonably well. A notable difference between the three methods is the sub-Poissonian statistics found in Fig.~\ref{fig3}(a), which is pronounced only in the case of averaging over the spherical shell. The reason is that the fringe pattern seen when plotting against $\phi$ washes out if the correlation function is averaged over a larger range of distances.

In summary, we discussed the second order correlation function for light emitted from a strongly and near-resonantly driven dilute cloud of atoms. Because of the strong driving, the fluorescence spectrum separates into distinct peaks, for which the spectral properties can be defined individually. We have shown that the second-order correlations for various combinations of photons from different spectral lines exhibit bunching together with super- or sub-Poissonian photon statistics, tunable by the choice of the detector positions. Furthermore, we demonstrated that the Cauchy-Schwarz inequality is violated for photons emitted from the two spectral side bands. The emitted light intensity is proportional to the square of the number of particles, and thus can potentially be intense. Interesting applications might arise if the presented method is applied to generate correlated photon pairs in the x-ray domain~\cite{mpik,bek,pal}.

\begin{acknowledgments}
The authors acknowledge valuable discussions with C. H. Keitel.
\end{acknowledgments}



\begin{thebibliography}{33}
\bibitem{RJG} R. J. Glauber, Rev. Mod. Phys. {\bf 78}, 1267 (2006);
L. Davidovich, {\it ibid} {\bf 68}, 127 (1996); H. Paul, {\it ibid}
{\bf 54}, 1061 (1982).

\bibitem{HBT1} R. Hanbury Brown, R. Q. Twiss, Nature (London) {\bf 177},
27 (1956).

\bibitem{mp} G. Baym, Acta Phys. Pol. B {\bf 29}, 1839 (1998).

\bibitem{HBT2} H. Hanbury Brown, R. Q. Twiss, Nature (London) {\bf 178},
1046 (1956).

\bibitem{ar} F. T. Arecchi, Phys. Rev. Lett. {\bf 15}, 912 (1965);
J. Beugnon, M. P. A. Jones, J. Dingjan, B. Darquie,
G. Messin, A. Browaeys, P. Grangier, Nature (London) {\bf 440}, 779
(2006).

\bibitem{hep} G. Goldhaber, S. Goldhaber, W. Lee, A. Pais,
Phys. Rev. {\bf 120}, 300 (1960).

\bibitem{cmp} M. Henny, S. Oberholzer, C. Strunk, T. Heinzel, K. Ensslin,
M. Holland, C. Sch\"{o}nenberger, Science {\bf 284}, 296 (1999);
W. D. Oliver, J. Kim, R. C. Liu, Y. Yamamoto, {\it ibid} {\bf 284}, 299 (1999).

\bibitem{dif} G. Sallen, A. Tribu, T. Aichele, R. Andre, L. Besombes, C. Bougerol,
M. Richard, S. Tatarenko, K. Kheng, J.-Ph. Poizat, Nature Photonics {\bf 4}, 696
(2010).

\bibitem{ap} M. Yasuda, F. Shimizu, Phys. Rev. Lett. {\bf 77}, 3090 (1996).

\bibitem{kl} D. N. Klyshko, {\it Photon and Nonlinear Optics}, (Gordon and
Breach Science Publishers, New York, 1989).

\bibitem{fwm1} S. Du, J. Wen, M. H. Rubin, G. Y. Yin, Phys. Rev. Lett.
{\bf 98}, 053601 (2007).

\bibitem{fwm2} M. Macovei, G.-x. Li, Phys. Rev. A {\bf 76}, 023818 (2007).

\bibitem{eit} V. Balic, D. A. Braje, P. Kolchin, G. Y. Yin,
S. E. Harris, Phys. Rev. Lett. {\bf 94}, 183601 (2005);
P. Kolchin, S. Du, C. Belthangady, G. Y. Yin, S. E.
Harris, {\it ibid} {\bf 97}, 113602 (2006).

\bibitem{ab_m} Ph. Grangier, G. Roger, A. Aspect, A. Heidmann,
S. Reynaud, Phys. Rev. Lett. {\bf 57}, 687 (1986).

\bibitem{kuz} A. Kuzmich, W. P. Bowen, A. D. Boozer, A. Boca,
C. W. Chou, L.-M. Duan, H. J. Kimble, Nature (London) {\bf 423}, 731 (2003).

\bibitem{semi} K. Edamatsu, G. Oohata, R. Shimizu, T. Itoh,
Nature (London) {\bf 431}, 167 (2004).

\bibitem{lk} C. H. van der Wal, M. D. Eisaman, A. Andre, R. L. Walsworth,
D. F. Phillips, A. S. Zibrov, M. D. Lukin, Science {\bf 301}, 196 (2003).

\bibitem{her} C. Wagenknecht, Che-M. Li, A. Reingruber, X.-H. Bao,
A. Goebel, Yu-Ao Chen, Q. Zhang, K. Chen, J.-W. Pan,
Nature Photonics {\bf 4}, 549 (2010); S. Barz, G. Cronenberg,
A. Zeilinger, Ph. Walther, {\it ibid} {\bf 4}, 553 (2010).


\bibitem{SP} W. H. Peeters, J. J. D. Moerman, M. P. van Exter,
Phys. Rev. Lett. {\bf 104}, 173601 (2010).

\bibitem{ks} T. Bienaime, S. Bux, E. Lucioni, P. W. Courteille,
N. Piovella, R. Kaiser, Phys. Rev. Lett. {\bf 104}, 183602 (2010).

\bibitem{fr} I. Shani, G. Cohen, J. Fineberg,
Phys. Rev. Lett. {\bf 104}, 184507 (2010).

\bibitem{scr} S. E. Skipetrov, R. Maynard,
Phys. Rev. Lett. {\bf 85}, 736 (2000).

\bibitem{carm} R. Carminati, Phys. Rev. A {\bf 81}, 053804 (2010).

\bibitem{ML} L. Sanchez-Palencia, M. Lewenstein,
Nature Phys. {\bf 6}, 87 (2010).


\bibitem{bks} G. Labeyrie, F. de Tomasi, J.-C. Bernard, C. A. M\"{u}ller, C. Miniatura,
R. Kaiser, Phys. Rev. Lett. {\bf 83}, 5266 (1999); V. Shatokhin, C. A. M\"{u}ller,
A. Buchleitner, {\it ibid} {\bf 94}, 043603 (2005).

\bibitem{rl} V. S. Letokhov, Sov. Phys. JETP {\bf 26}, 835 (1968).

\bibitem{rl1} M. Patra, Phys. Rev. A {\bf 65}, 043809 (2002).

\bibitem{rl2} L. S. Froufe-Perez, W. Guerin, R. Carminati, R. Kaiser,
Phys. Rev. Lett. {\bf 102}, 173903 (2009); W. Guerin, F. Michaud, R.
Kaiser, {\it ibid} {\bf 101}, 093002 (2008).

\bibitem{melk} M. Macovei, J. Evers, G. X. Li, C. H. Keitel,
Phys. Rev. Lett. {\bf 98}, 043602 (2007); M. Macovei, J. Evers,
C. H. Keitel, arXiv:quant-ph/0702142v1.

\bibitem{Mollow} B. R. Mollow, Phys. Rev. {\bf 188}, 1969 (1969).

\bibitem{kilin} P. A. Apanasevich, S. J. Kilin, J. Phys. B: At. Mol.
Phys. {\bf 12}, L83 (1979).

\bibitem{Tann} C. Cohen-Tannoudji, R. Reynaud,
Phil. Trans. R. Soc. Lond. A {\bf 293}, 223 (1979).

\bibitem{shat} V. Shatokhin, C. A. M\"{u}ller,
A. Buchleitner, Phys. Rev. A {\bf 73}, 063813 (2006).

\bibitem{CSI} J. F. Clauser, Phys. Rev. D {\bf 9}, 853 (1974); R. Loudon, Rep.
Prog. Phys. {\bf 43}, 58 (1980).

\bibitem{mpik} S. W. Epp, J. R. Crespo Lopez-Urrutia, G. Brenner, V. M\"{a}ckel,
P. H. Mokler, R. Treusch, M. Kuhlmann, M. V. Yurkov, J. Feldhaus, J. R. Schneider,
M. Wellh\"{o}fer, M. Martins, W. Wurth, J. Ullrich, Phys. Rev. Lett. {\bf 98}, 183001
(2007).

\bibitem{bek} Th. B\"{u}rvenich, J. Evers, and C. H. Keitel, Phys. Rev. Lett. {\bf 96},
142501 (2006).

\bibitem{pal}A. P\'alffy, C. H. Keitel and J. Evers, Phys. Rev. Lett. {\bf 103}, 017401 (2009)

\end{thebibliography}
\end{document}